\documentclass[aps,twocolumn]{revtex4}

\usepackage{graphicx}

\newcommand{\myfigsize}{0.9}

\newcommand{\myd}[2]{\frac{\partial #1}{\partial #2}}

\newcommand{\mysum}[2]{\sum\limits_{#1}^{#2}}
\newcommand{\myint}[2]{\int\limits_{#1}^{#2}}

\newcommand{\mybpar}[1]{\left( #1 \right)}
\newcommand{\beq}{\begin{equation}}
\newcommand{\eeq}{\end{equation}}
\newcommand{\beqa}{\begin{eqnarray}}
\newcommand{\eeqa}{\end{eqnarray}}

\begin{document}
\title{Thermoelectric effect in molecular electronics} 
\author{Magnus Paulsson}
\surname{Paulsson}
\email{mpaulsso@purdue.edu}
\author{Supriyo Datta}
\surname{Datta}
\email{datta@purdue.edu}
\affiliation{
Purdue University, School Of Electrical \& Computer Engineering,  \\
1285 Electrical Engineering Building, 
West Lafayette, IN - 47907-1285, USA\\
www.nanohub.purdue.edu} 
\date{\today} 

\begin{abstract}
We provide a theoretical estimate of the thermoelectric current and voltage over a Phenyldithiol molecule. 
We also show that the thermoelectric voltage is (1) easy to analyze, (2) insensitive to the detailed 
coupling to the contacts, (3) large enough to be measured and (4) give valuable information, which is not 
readily accessible through other experiments, on the location of the Fermi energy relative to the molecular levels. 
The location of the Fermi-energy is poorly understood and controversial even though it is a central factor 
in determining the nature of conduction (n- or p-type). We also note that the thermoelectric voltage 
measured over Guanine molecules with an STM by Poler {\it et al.}, indicate conduction through 
the HOMO level, i.e., p-type conduction. 
\end{abstract}


\maketitle

\section{Introduction}
Electrical conduction through individual molecules chemically bound to gold contacts has recently been measured using STM \cite{donhauser.s01,datta.sup00}, break junctions \cite{weber.prl02} and nanopores \cite{reed.cp02}. Theoretical calculations of the current-voltage (I-V) characteristics have also been reported, see for example Ref. \cite{damle.cp02,stokbro.cms03,lang.prl00,palacios.prb02}. However, a detailed quantitative comparison of theoretical and experimental results is made difficult by two factors. Firstly, the low bias conductance depends strongly on the quality of the metal-molecule contacts \cite{stokbro.cms03}, which is ill controlled and poorly characterized experimentally. Secondly, the conductance gap is determined by the location of the Fermi energy relative to the molecular levels - a factor that is poorly understood and controversial \cite{paulsson.crc02,ghosh.asym02}. This is particularly unfortunate since the location of the Fermi energy is a central factor in determining the nature of conduction (n- or p-type).

\begin{figure}
\begin{center}
\includegraphics[width=\myfigsize \columnwidth]{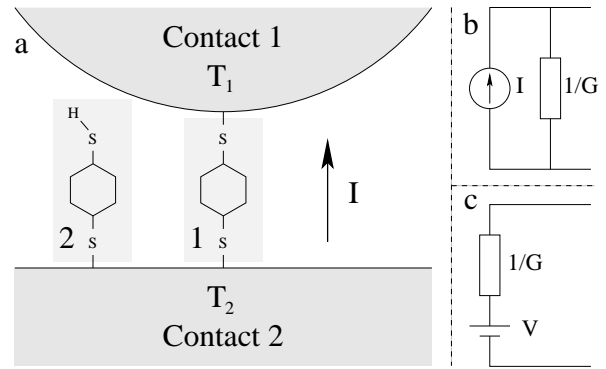}
\end{center}
\caption{a: Proposed experimental setup to measure thermoelectric effects over a molecule 
with two contacts  at different temperatures $T_1$, $T_2$. 
In this paper we focus on the Phenyl-dithiol (PDT) molecule (1) chemically bound to 
both contacts (break junction setup) and (2) with one weak contact (STM measurement).  
b, c: Equivalent circuits defining the voltage, conductance and current.}
\label{fig.overview}
\end{figure}

The purpose of this paper is to show that a measurement of the thermoelectric voltage (see Fig. \ref{fig.overview}) can provide new insights into electron transport and allows us to estimate the location of the Fermi energy relative to the molecular levels. Analogous to the hot point probe measurements commonly used to establish the p- or n-character of semi-conductors (see for example Ref. \cite{pierret.thermo}), the thermoelectric voltage yield valuable information regarding the location of the Fermi energy. In contrast to the I-V, the thermoelectric voltage is an equilibrium property which is easy to interpret. What makes it particularly useful is that unlike the low-bias conductance, it is relatively insensitive to the quality of the contacts.

In this paper we estimate the thermoelectric current and voltage over a Phenyldithiol (PDT) molecule, see Fig. \ref{fig.overview}. We show that the thermoelectric voltage is (1) easy to analyze, (2) insensitive to the contacts and (3) large enough to be measured. In fact, Poler {\it et al.} \cite{poler.langmuir95} measured the thermoelectric voltage (with a $20$ K temperature difference) over a monolayer of Guanine molecules with an STM. In this paper we also show that the thermoelectric voltage provide information on where the Fermi energy is relative to the molecular levels, e.g., the sign of the thermoelectric voltage over the Guanine molecule indicate p-type conduction. 

\section{Method}
The electrical transport through small conjugated molecules chemically bound to at least one of the contacts can be described by the Landauer formula in terms of the transmission ($\mathcal{T}(E)$) \cite{buttiker.prb85}:
\beq
I=\frac{e}{\pi \hbar} \myint{-\infty}{\infty}
{\mathcal{T}(E) \mybpar{f_1(E)-f_2(E)} \; \mbox{d}E}
\eeq
assuming that the electronic states in the contacts are filled according to
the Fermi distribution of the reservoirs ($f_1$, $f_2$).
If the transmission is not constant (as a function of energy), a difference in the Fermi distributions due to a 
temperature difference ($\Delta T=T_1-T_2$) will drive a thermoelectric current at zero bias ($V=0$)
(also see footnote \footnote{In addition to the current given by the slope of the transmission, the 
chemical potentials of the contacts shifts slightly. However, this effect
is negligible for most metals \cite{ashcroft.chempot}.}). 
For molecules the transmission is often smooth (compared to the thermal energy) which allows us to 
use the Sommerfeld expansion \cite{butcher.jpc90}:
\beq
I=-\frac{e^2}{\pi \hbar} \mathcal{T}(E_f) V +  \frac{e}{\pi \hbar}\frac{\pi^2 k_B^2 T}{3}\left. 
\myd{\mathcal{T}(E)}{\mbox{E}}\right|_{E=E_f} \Delta T 
\label{eq.current}
\eeq
where $T$ is the mean temperature of the contacts ($(T_1+T_2)/2$). 
Similar expressions are available for heat transport \cite{butcher.jpc90}.

The thermoelectric current is usually small enough that we can use a linear equivalent circuit as shown in Fig. \ref{fig.overview} b.
The current source (see Eq. \ref{eq.current}) has a resistance that
is the inverse of the low bias conductance:
\beq
G=\frac{e^2}{\pi \hbar} \mathcal{T}(E_f)
\label{eq.conductance}
\eeq
Alternatively we could represent the effect in terms of a voltage source ($V=I/G$) with a series resistance as shown in Fig. \ref{fig.overview} c:
\beq
 \left. V \right|_{I=0}=\frac{\pi^2 k_B^2 T}{3 e} 
 {\left.\myd{\ln\mybpar{\mathcal{T}(E)}}{E}\right|_{E=E_f}} \Delta T
\label{eq.voltage}
\eeq
As we will see below, this open circuit voltage can be large enough to measure
and is relatively insensitive to the coupling to the contacts.

To obtain estimates of the low bias conductance and the thermoelectric voltage 
we need to calculate the transmission as a function of energy for a molecule connected to
metallic contacts. Here we calculate the transmission using the Non Equilibrium 
Green's Function (NEGF) formalism using Extended H\"uckel (EH) theory (see footnote 
\footnote{To compare energy levels with experimental quantities, e.g., work-functions, 
the EH levels should be shifted (towards positive energies) by approximately $4.6$ 
eV since electron-electron interactions are neglected.})
as described in Ref. \cite{zahid.mark02_nanohub}. 
This is a simple but sufficient approximation if we take $E_f$ as an adjustable parameter
(see footnote \footnote{In reality the Fermi-energy of the contacts is fixed but 
charging effects shift the molecular levels. Here we keep the molecular levels fixed 
and treat the Fermi-energy of the contacts as a parameter.}) 
to be determined from experiments since the qualitative features of the transmission, 
in our experience, is not very sensitive to the method (H\"uckel, EH or Ab-initio) used. 

\section{An illustrative example}
We choose to estimate the thermoelectric voltage across a phenyl-dithiol (PDT) molecule 
(Fig. \ref{fig.overview}a) since it 
has been studied extensively after the experiment by Reed {\it et al.} \cite{reed.s97}.
In any case, our objective is to provide a reasonable qualitative estimate rather 
than an accurate quantitative value for a specific molecule.  
On gold surfaces, PDT chemisorbs forming a bond between the sulfur and gold.
The strength of this bond is difficult to estimate since the precise experimental geometry is unknown. 
For an STM measurement the contact with the tip is also weaker than the substrate bond. 
We therefore calculate the transmission through the PDT molecule perfectly bound to the 
gold contacts (see molecule 1 in Fig. \ref{fig.overview} a) and through a PDT molecule 
chemisorbed on only one contact with different distances between the molecule and 
contact 1 (see molecule 2 in Fig. \ref{fig.overview}a). 
 
\begin{figure}[!hbt]
\begin{center}
\includegraphics[width=\myfigsize \columnwidth]{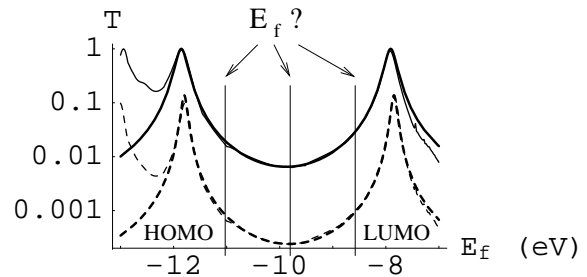}
\end{center}
\caption{Logarithmic plot of the transmission as a function of energy.
Thin solid line: Transmission calculated from the Extended H\"uckel model through 
the PDT molecule strongly bound to both contacts. 
Thin dashed line: PDT weakly bound to one side (Au-H distance 2.9\AA).
Thick lines: The transmission fitted to Lorentzian transmission peaks associated with the 
molecular energy levels (see Eq. \ref{eq.trans}).
Note the widely different positions of the Fermi energy ($E_f$) suggested by different authors.  
}
\label{fig.PDT_T}
\end{figure}

Fig. \ref{fig.PDT_T} shows the transmission through the PDT molecule for 
perfect contacts on both sides (solid line) and one weak contact (dashed line). 
The position of the Fermi energy ($E_f$) relative to the Highest Occupied Molecular Orbital (HOMO) 
and Lowest Unoccupied Molecular Orbital (LUMO) levels is one of the main
parameters affecting the current-voltage characteristics \cite{paulsson.crc02}. 
As pointed out in the introduction, the position of $E_f$ 
is difficult to estimate.
The suggested positions of $E_f$ range 
from (1) closer to the HOMO level \cite{damle.cp02,emberly.prb98,ratner.jcp99,stokbro.cms03},
(2) mid gap and (3) closer to the LUMO level \cite{lang.prl00}, see Fig. \ref{fig.PDT_T}.
Since the thermoelectric effect depends on the slope of the transmission at 
$E_f$ (Eq. \ref{eq.voltage}), the sign of the thermoelectric voltage and 
current will be determined by where $E_f$ is located relative to the molecular levels. 

\begin{figure}[!hbt]
\begin{center}
\includegraphics[width=\myfigsize \columnwidth]{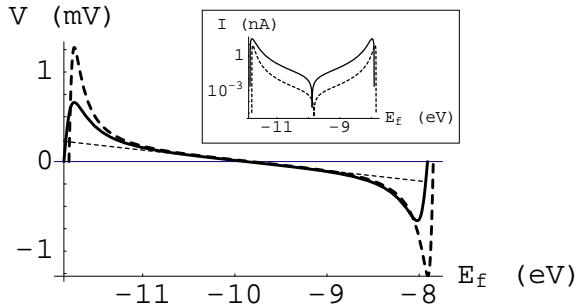}
\end{center}
\caption{Thermoelectric voltage generated by a $10$ K difference in temperature 
($T_1=300$ K, $T_2=310$ K) for different positions of the Fermi energy ($E_f$). 
Solid line: PDT strongly bound to both contacts. Dashed line: weak bond on one side.
Thin dotted line, approximated by Eq. \ref{eq.approxV}. Inset: Logarithmic plot of the thermoelectric current.}.
\label{fig.PDT_V}
\end{figure}

The thermoelectric voltage calculated from Eqs. \ref{eq.voltage}, \ref{eq.trans} is shown in 
Fig. \ref{fig.PDT_V} for a $10$ K temperature difference at room temperature 
($T_1=300$ K, $T_2=310$ K). 
Note that the thermoelectric voltage was calculated from a transmission fitted to Lorentzian peaks 
(see Fig. \ref{fig.PDT_T}), this removes the magnification (by taking the derivative) of small 
numerical errors.
Unless $E_f$ is located very close to the middle of the HOMO-LUMO gap the size of the 
thermoelectric voltage is of the order $~0.5-0.1$ mV. 
Since the thermoelectric voltage is proportional to $\Delta T$ (Eq. \ref{eq.voltage}), 
even smaller temperature differences will give a measurable thermoelectric voltage.

The measured conductance of the PDT molecules \cite{reed.s97} 
is smaller than the theoretical estimates by two to three orders of magnitude \cite{stokbro.cms03}. 
This discrepancy indicate that the interaction strength between molecule and contacts
is smaller than that assumed in the models (for STM measurements this is obviously true). 
It is plausible to believe that the molecule form
a strong bond with one of the contacts and only interacts weakly with the second contact
(also see footnote \footnote{Emberly and Kirczenow have pointed out that the 
symmetric I-V curves obtained in
measurements on PDT are consistent with a geometry where two weakly interacting molecules are 
attach to the two different contacts \cite{emberly.prl01}. We have omitted this case in this paper
since the thermoelectric voltage calculated in this geometry is qualitatively the same as for one 
PDT molecule weakly coupled to one contact.}). 
However, the calculated thermoelectric voltage with one weak contact 
(Fig. \ref{fig.PDT_V} dashed line) 
has almost the same magnitude as the perfectly bound molecule 
(solid line) and is slightly larger close to the transmission peaks.

To understand why the thermoelectric voltage is unaffected by weak contacts it is instructive to 
fit the transmission (Fig. \ref{fig.PDT_T}) to Lorentzian transmission peaks:
\beq
\mathcal{T}(E)=\mysum{i=1}{2}
\frac{\Gamma_1 \Gamma_2}{\mybpar{E-\epsilon_i}^2+\mybpar{\Gamma_1+\Gamma_2}^2/4}
\label{eq.trans}
\eeq
where $\epsilon_i$ is the energy of the two levels and $\Gamma_{1}, \Gamma_2$ the
broadening by contact $1$ and $2$.
This approximation is useful for the PDT molecule since the transmission around the 
HOMO-LUMO gap of PDT is mainly determined by two levels that 
couple approximately equally to the contacts (see footnote \footnote{The transmission peaks 
actually correspond to the HOMO and LUMO+1 levels since the LUMO level is localized to the 
benzene ring and give zero transmission.}). Fitting the transmission
to Eq. \ref{eq.trans} using a least square fit (of $\ln(\mathcal{T})$) gives:
$\Gamma_1 = \Gamma_2 = 0.11$ eV, $\epsilon_1=-11.86$ eV and $\epsilon_2=-7.91$ eV. 
For the weakly coupled PDT molecule (with a Au-H distance of $2.9$ \AA) 
the least squares fit gives ($\Gamma_2 = 0.11$ eV): $\Gamma_1 = 0.0042$ eV, 
$\epsilon_1=-11.80$ eV, $\epsilon_2=-7.85$ eV. These fits give good approximations 
to the calculated transmission (see Fig. \ref{fig.PDT_T}), other Au-H distances 
(not shown in figure) in the range $1-5$ \AA \ also fit the transmission well.

Assuming the Fermi energy to be situated in between the HOMO and LUMO levels and far 
from the levels ($|E_f-\epsilon_{1,2}| >> |\Gamma_1+\Gamma_2|$) 
we can Taylor expand Eq. \ref{eq.voltage} in the energy around 
the midpoint of the HOMO-LUMO gap ($E=\frac{\epsilon_1+\epsilon_2}{2}$):
\beq
\left.V\right|_{I=0} =\frac{8 \pi^2 k_B^2 T}{e}\frac{1}{\mybpar{\epsilon_1-\epsilon_2}^2} 
\mybpar{E_f-\frac{\epsilon_1+\epsilon_2}{2}} \Delta T\label{eq.approxV}
\eeq
This shows that the thermoelectric voltage is, to first order, independent of 
the strength of the interaction with the contacts, see thin dotted line in Fig. \ref{fig.PDT_V}. 
This result is not limited to the PDT molecule
but valid for most conjugated molecules (see footnote \footnote{
From the NEGF formalism \cite{paulsson.crc02}, $\mathcal{T}=
\mbox{Tr}(G \Gamma_1 G^\dagger \Gamma_2)$.
If a weak coupling only scales $\Gamma_1$ by a constant, it is easy to show that the transmission 
scales by the same constant when the energy is far from any resonances (due to the molecular 
energy levels) in $G$.}). 

To experimentally measure the thermoelectric voltage, the thermoelectric current has to be larger 
than any leakage currents.  
The thermoelectric current (Eq. \ref{eq.current}) through the PDT molecule is shown in the inset of 
Fig. \ref{fig.PDT_V}. 
Since the thermoelectric current is proportional to the transmission ($\mathcal{T}$), 
the magnitude of the current is strongly dependent on the strength of the interactions 
between molecule and contacts. It is also important to realize that for very weak coupling 
to both contacts, i.e., in the Coulomb blockade regime, the simple decomposition of the transmission into
HOMO and LUMO peaks (see Fig. \ref{fig.PDT_T}) might not be enough to describe the thermoelectric effect. 
However, even for a molecule which is relatively weakly connected to one contact 
(but chemically bound to the other contact) should give a thermoelectric current of the order of $1-100$ pA, 
see inset in Fig. \ref{fig.PDT_V}. 

\section{Summary and discussion}
\label{sect.summary}

The position of the Fermi energy is one important ingredient in understanding electrical transport through molecules. 
It is also interesting since shifting the Fermi energy by, (1) doping, (2) addition of side-groups, (3) gate field or (4) use of different contact material, can be used to optimize the molecular properties. Phenomena of this type can be probed and evaluated through a measurement of the thermoelectric voltage (coupled with I-V measurements).

In this study we have only considered a single molecule between the two contacts. However, unless the local structure of each molecule is different, the thermoelectric effect should only be affected trivially. It is possible that the detailed structure of the gold atoms bonding to the molecule affect the charge transfer to the molecule and thus the position of the Fermi energy. If this is the case, the variations could be mapped by STM measurements on different molecules on a single substrate. Another simplification in our work is that we only focus on the PDT molecule. Due to its simple structure the transmission is almost symmetric around the middle of the HOMO-LUMO gap. For other molecules this need not be true. In this case, the interpretation of the thermoelectric voltage still provides an estimate of the Fermi energy even though it is less straight forward. 

To our knowledge the only reported measurement of the thermoelectric voltage over a molecule was performed by Poler {\it et al.} \cite{poler.langmuir95}. Using an STM tip they measured the thermoelectric voltage over Guanine molecules on a graphite substrate. A detailed analysis of the measurement is not possible since the I-V characteristic was not measured in this experiment.  However, the sign of the thermoelectric voltage indicates that the electrical transport is conducted through the HOMO level and the measured thermoelectric voltage was similar in magnitude to the value estimated here for PDT ($~0.5\pm0.01$ mV at $\Delta T=26.5$ K). 

An intriguing possibility is to use the thermoelectric effect over a molecular SAM as a thermoelectric element. Applications of these elements include thermoelectric coolers (Peltier effect) and power generators. We have performed preliminary calculations on the efficiency of such elements. However, the efficiency is limited (and largely determined) by the lattice thermal conductivities of the SAM. We therefore believe that until there are reliable estimates of the thermal conductivity any speculations on the efficiency of such elements are premature.

In summary, we have shown that a measurement of the thermoelectric voltage over a molecule is (1) simple to analyze, (2) insensitive to the contacts, (3) feasible and (4) should give valuable information, which is not readily accessible through other experiments, on the location of the Fermi energy relative to the molecular levels. 


\begin{thebibliography}{20}
\expandafter\ifx\csname natexlab\endcsname\relax\def\natexlab#1{#1}\fi
\expandafter\ifx\csname bibnamefont\endcsname\relax
  \def\bibnamefont#1{#1}\fi
\expandafter\ifx\csname bibfnamefont\endcsname\relax
  \def\bibfnamefont#1{#1}\fi
\expandafter\ifx\csname citenamefont\endcsname\relax
  \def\citenamefont#1{#1}\fi
\expandafter\ifx\csname url\endcsname\relax
  \def\url#1{\texttt{#1}}\fi
\expandafter\ifx\csname urlprefix\endcsname\relax\def\urlprefix{URL }\fi
\providecommand{\bibinfo}[2]{#2}
\providecommand{\eprint}[2][]{\url{#2}}

\bibitem[{\citenamefont{Hong et~al.}(2000)\citenamefont{Hong, Reifenberger,
  Tian, Datta, Henderson, and Kubiak}}]{datta.sup00}
\bibinfo{author}{\bibfnamefont{S.}~\bibnamefont{Hong}},
  \bibinfo{author}{\bibfnamefont{R.}~\bibnamefont{Reifenberger}},
  \bibinfo{author}{\bibfnamefont{W.}~\bibnamefont{Tian}},
  \bibinfo{author}{\bibfnamefont{S.}~\bibnamefont{Datta}},
  \bibinfo{author}{\bibfnamefont{J.}~\bibnamefont{Henderson}},
  \bibnamefont{and} \bibinfo{author}{\bibfnamefont{C.~P.}
  \bibnamefont{Kubiak}}, \bibinfo{journal}{Superlattices and Microstructures}
  \textbf{\bibinfo{volume}{28}}, \bibinfo{pages}{289} (\bibinfo{year}{2000}).

\bibitem[{\citenamefont{Donhauser et~al.}(2001)\citenamefont{Donhauser,
  Mantooth, Kelly, Bumm, Monnell, Stapleton, Price, Rawlett, Allara, Tour
  et~al.}}]{donhauser.s01}
\bibinfo{author}{\bibfnamefont{Z.~J.} \bibnamefont{Donhauser}},
  \bibinfo{author}{\bibfnamefont{B.~A.} \bibnamefont{Mantooth}},
  \bibinfo{author}{\bibfnamefont{K.~F.} \bibnamefont{Kelly}},
  \bibinfo{author}{\bibfnamefont{L.~A.} \bibnamefont{Bumm}},
  \bibinfo{author}{\bibfnamefont{J.~D.} \bibnamefont{Monnell}},
  \bibinfo{author}{\bibfnamefont{J.~J.} \bibnamefont{Stapleton}},
  \bibinfo{author}{\bibfnamefont{D.~W.} \bibnamefont{Price}},
  \bibinfo{author}{\bibfnamefont{A.~M.} \bibnamefont{Rawlett}},
  \bibinfo{author}{\bibfnamefont{D.~L.} \bibnamefont{Allara}},
  \bibinfo{author}{\bibfnamefont{J.~M.} \bibnamefont{Tour}},
  \bibnamefont{et~al.}, \bibinfo{journal}{Science}
  \textbf{\bibinfo{volume}{292}}, \bibinfo{pages}{2303} (\bibinfo{year}{2001}).

\bibitem[{\citenamefont{Reichert et~al.}(2002)\citenamefont{Reichert, Ochs,
  Beckmann, Weber, Mayor, and L\"ohneysen}}]{weber.prl02}
\bibinfo{author}{\bibfnamefont{J.}~\bibnamefont{Reichert}},
  \bibinfo{author}{\bibfnamefont{R.}~\bibnamefont{Ochs}},
  \bibinfo{author}{\bibfnamefont{D.}~\bibnamefont{Beckmann}},
  \bibinfo{author}{\bibfnamefont{H.~B.} \bibnamefont{Weber}},
  \bibinfo{author}{\bibfnamefont{M.}~\bibnamefont{Mayor}}, \bibnamefont{and}
  \bibinfo{author}{\bibfnamefont{H.~v.} \bibnamefont{L\"ohneysen}},
  \bibinfo{journal}{Phys. Rev. Lett.} \textbf{\bibinfo{volume}{88}},
  \bibinfo{pages}{176804} (\bibinfo{year}{2002}).

\bibitem[{\citenamefont{Chen and Reed}(2002)}]{reed.cp02}
\bibinfo{author}{\bibfnamefont{J.}~\bibnamefont{Chen}} \bibnamefont{and}
  \bibinfo{author}{\bibfnamefont{M.~A.} \bibnamefont{Reed}},
  \bibinfo{journal}{Chem. Phys.}  (\bibinfo{year}{2002}).

\bibitem[{\citenamefont{Damle et~al.}(2002)\citenamefont{Damle, Ghosh, and
  Datta}}]{damle.cp02}
\bibinfo{author}{\bibfnamefont{P.~S.} \bibnamefont{Damle}},
  \bibinfo{author}{\bibfnamefont{A.~W.} \bibnamefont{Ghosh}}, \bibnamefont{and}
  \bibinfo{author}{\bibfnamefont{S.}~\bibnamefont{Datta}},
  \bibinfo{journal}{Chem. Phys.} \textbf{\bibinfo{volume}{281}},
  \bibinfo{pages}{171} (\bibinfo{year}{2002}).

\bibitem[{\citenamefont{Di~Ventra et~al.}(2000)\citenamefont{Di~Ventra,
  Pantelides, and Lang}}]{lang.prl00}
\bibinfo{author}{\bibfnamefont{M.}~\bibnamefont{Di~Ventra}},
  \bibinfo{author}{\bibfnamefont{S.~T.} \bibnamefont{Pantelides}},
  \bibnamefont{and} \bibinfo{author}{\bibfnamefont{N.~D.} \bibnamefont{Lang}},
  \bibinfo{journal}{Phys. Rev. Lett.} \textbf{\bibinfo{volume}{84}},
  \bibinfo{pages}{979} (\bibinfo{year}{2000}).

\bibitem[{\citenamefont{Palacios et~al.}(2002)\citenamefont{Palacios, Louis,
  SanFabi\'an, and Verg\'es}}]{palacios.prb02}
\bibinfo{author}{\bibfnamefont{A.~J.} \bibnamefont{Palacios},
  \bibfnamefont{J.~J.and P\'erez-Jim\'enez}},
  \bibinfo{author}{\bibfnamefont{E.}~\bibnamefont{Louis}},
  \bibinfo{author}{\bibfnamefont{E.}~\bibnamefont{SanFabi\'an}},
  \bibnamefont{and} \bibinfo{author}{\bibfnamefont{J.~A.}
  \bibnamefont{Verg\'es}}, \bibinfo{journal}{Phys. Rev. B}
  \textbf{\bibinfo{volume}{66}}, \bibinfo{pages}{035322}
  (\bibinfo{year}{2002}).

\bibitem[{\citenamefont{Stokbro et~al.}(2003)\citenamefont{Stokbro, Taylor,
  Brandbyge, Mozos, and Ordej\'{o}n}}]{stokbro.cms03}
\bibinfo{author}{\bibfnamefont{K.}~\bibnamefont{Stokbro}},
  \bibinfo{author}{\bibfnamefont{J.}~\bibnamefont{Taylor}},
  \bibinfo{author}{\bibfnamefont{M.}~\bibnamefont{Brandbyge}},
  \bibinfo{author}{\bibfnamefont{J.-L.} \bibnamefont{Mozos}}, \bibnamefont{and}
  \bibinfo{author}{\bibfnamefont{P.}~\bibnamefont{Ordej\'{o}n}},
  \bibinfo{journal}{Computational materials science}  (\bibinfo{year}{2003}),
  \bibinfo{note}{in press}.

\bibitem[{\citenamefont{Paulsson et~al.}(2002)\citenamefont{Paulsson, Zahid,
  and Datta}}]{paulsson.crc02}
\bibinfo{author}{\bibfnamefont{M.}~\bibnamefont{Paulsson}},
  \bibinfo{author}{\bibfnamefont{F.}~\bibnamefont{Zahid}}, \bibnamefont{and}
  \bibinfo{author}{\bibfnamefont{S.}~\bibnamefont{Datta}},
  \emph{\bibinfo{title}{Nanoscience, Engineering and Technology Handbook}}
  (\bibinfo{publisher}{CRC Press}, \bibinfo{year}{2002}), chap.
  \bibinfo{chapter}{Resistance of a Molecule}, \bibinfo{note}{editors D.
  Brenner, S. Lyshevski and G. Iafrate, cond-mat/0208183.}

\bibitem[{\citenamefont{Ghosh et~al.}()\citenamefont{Ghosh, Zahid, Damle, and
  Datta}}]{ghosh.asym02}
\bibinfo{author}{\bibfnamefont{A.~W.} \bibnamefont{Ghosh}},
  \bibinfo{author}{\bibfnamefont{F.}~\bibnamefont{Zahid}},
  \bibinfo{author}{\bibfnamefont{P.~S.} \bibnamefont{Damle}}, \bibnamefont{and}
  \bibinfo{author}{\bibfnamefont{S.}~\bibnamefont{Datta}},
  \bibinfo{note}{cond-mat/0202519}.

\bibitem[{\citenamefont{Pierret}(1996)}]{pierret.thermo}
\bibinfo{author}{\bibfnamefont{R.~F.} \bibnamefont{Pierret}},
  \emph{\bibinfo{title}{Semiconductor Fundamentals}}
  (\bibinfo{publisher}{Addison-Wesley}, \bibinfo{year}{1996}), chap.
  \bibinfo{chapter}{3.2.2}.

\bibitem[{\citenamefont{Poler et~al.}(1995)\citenamefont{Poler, Zimmermann, and
  Cox}}]{poler.langmuir95}
\bibinfo{author}{\bibfnamefont{J.~C.} \bibnamefont{Poler}},
  \bibinfo{author}{\bibfnamefont{R.~M.} \bibnamefont{Zimmermann}},
  \bibnamefont{and} \bibinfo{author}{\bibfnamefont{E.~C.} \bibnamefont{Cox}},
  \bibinfo{journal}{Langmuir} \textbf{\bibinfo{volume}{11}},
  \bibinfo{pages}{2689} (\bibinfo{year}{1995}).

\bibitem[{\citenamefont{B\"{u}ttiker et~al.}(1985)\citenamefont{B\"{u}ttiker,
  Imry, Landauer, and Pinhas}}]{buttiker.prb85}
\bibinfo{author}{\bibfnamefont{M.}~\bibnamefont{B\"{u}ttiker}},
  \bibinfo{author}{\bibfnamefont{Y.}~\bibnamefont{Imry}},
  \bibinfo{author}{\bibfnamefont{R.}~\bibnamefont{Landauer}}, \bibnamefont{and}
  \bibinfo{author}{\bibfnamefont{S.}~\bibnamefont{Pinhas}},
  \bibinfo{journal}{Phys. Rev. B} \textbf{\bibinfo{volume}{31}},
  \bibinfo{pages}{6207} (\bibinfo{year}{1985}).

\bibitem[{\citenamefont{Butcher}(1990)}]{butcher.jpc90}
\bibinfo{author}{\bibfnamefont{P.~N.} \bibnamefont{Butcher}},
  \bibinfo{journal}{J. Phys. Chem.} \textbf{\bibinfo{volume}{2}},
  \bibinfo{pages}{4869} (\bibinfo{year}{1990}).

\bibitem[{\citenamefont{Zahid et~al.}(2002)\citenamefont{Zahid, Paulsson, and
  Datta}}]{zahid.mark02_nanohub}
\bibinfo{author}{\bibfnamefont{F.}~\bibnamefont{Zahid}},
  \bibinfo{author}{\bibfnamefont{M.}~\bibnamefont{Paulsson}}, \bibnamefont{and}
  \bibinfo{author}{\bibfnamefont{S.}~\bibnamefont{Datta}},
  \emph{\bibinfo{title}{Advanced Semiconductors and Organic Nano-Techniques}}
  (\bibinfo{publisher}{Academic press}, \bibinfo{year}{2002}), chap.
  \bibinfo{chapter}{Electrical Conduction through Molecules},
  \bibinfo{note}{editor H. Markoc, to be published in 2002, the program
  (Huckel-IV 2.0) used to calculate the transmission and the preprint are
  availible at 'www.nanohub.purdue.edu'.}

\bibitem[{\citenamefont{Reed et~al.}(1997)\citenamefont{Reed, Zhou, Muller,
  Burgin, and Tour}}]{reed.s97}
\bibinfo{author}{\bibfnamefont{M.~A.} \bibnamefont{Reed}},
  \bibinfo{author}{\bibfnamefont{C.}~\bibnamefont{Zhou}},
  \bibinfo{author}{\bibfnamefont{C.~J.} \bibnamefont{Muller}},
  \bibinfo{author}{\bibfnamefont{T.~P.} \bibnamefont{Burgin}},
  \bibnamefont{and} \bibinfo{author}{\bibfnamefont{J.~M.} \bibnamefont{Tour}},
  \bibinfo{journal}{Science} \textbf{\bibinfo{volume}{278}},
  \bibinfo{pages}{252} (\bibinfo{year}{1997}).

\bibitem[{\citenamefont{Emberly and Kirczenow}(1998)}]{emberly.prb98}
\bibinfo{author}{\bibfnamefont{E.~G.} \bibnamefont{Emberly}} \bibnamefont{and}
  \bibinfo{author}{\bibfnamefont{G.}~\bibnamefont{Kirczenow}},
  \bibinfo{journal}{Phys. Rev. B} \textbf{\bibinfo{volume}{58}},
  \bibinfo{pages}{10911} (\bibinfo{year}{1998}).

\bibitem[{\citenamefont{Yaliraki et~al.}(1999)\citenamefont{Yaliraki, Roitberg,
  Gonzalez, Mujica, and Ratner}}]{ratner.jcp99}
\bibinfo{author}{\bibfnamefont{S.~N.} \bibnamefont{Yaliraki}},
  \bibinfo{author}{\bibfnamefont{A.~E.} \bibnamefont{Roitberg}},
  \bibinfo{author}{\bibfnamefont{C.}~\bibnamefont{Gonzalez}},
  \bibinfo{author}{\bibfnamefont{V.}~\bibnamefont{Mujica}}, \bibnamefont{and}
  \bibinfo{author}{\bibfnamefont{M.~A.} \bibnamefont{Ratner}},
  \bibinfo{journal}{J. Chem. Phys.} \textbf{\bibinfo{volume}{111}},
  \bibinfo{pages}{6997} (\bibinfo{year}{1999}).

\bibitem[{\citenamefont{Ashcroft and Mermin}(1976)}]{ashcroft.chempot}
\bibinfo{author}{\bibfnamefont{N.~W.} \bibnamefont{Ashcroft}} \bibnamefont{and}
  \bibinfo{author}{\bibfnamefont{N.~D.} \bibnamefont{Mermin}},
  \emph{\bibinfo{title}{Solid state physics}} (\bibinfo{publisher}{{Saunders
  college publishing}}, \bibinfo{address}{Philadelphia}, \bibinfo{year}{1976}),
  chap.~\bibinfo{chapter}{2}.

\bibitem[{\citenamefont{Emberly and Kirczenow}(2001)}]{emberly.prl01}
\bibinfo{author}{\bibfnamefont{E.~G.} \bibnamefont{Emberly}} \bibnamefont{and}
  \bibinfo{author}{\bibfnamefont{G.}~\bibnamefont{Kirczenow}},
  \bibinfo{journal}{Phys. Rev. Lett.} \textbf{\bibinfo{volume}{87}}
  (\bibinfo{year}{2001}).

\end{thebibliography}

\end{document}